\documentclass[%
 aip,
 amsmath,amssymb,
 reprint,%
]{revtex4-1}

\usepackage{graphicx}
\usepackage{dcolumn}
\usepackage{bm}

\usepackage[utf8]{inputenc}
\usepackage[T1]{fontenc}
\usepackage{mathptmx}
\usepackage{etoolbox}
\usepackage{amsthm,amsmath}

\makeatletter
\def\@email#1#2{%
 \endgroup
 \patchcmd{\titleblock@produce}
  {\frontmatter@RRAPformat}
  {\frontmatter@RRAPformat{\produce@RRAP{*#1\href{mailto:#2}{#2}}}\frontmatter@RRAPformat}
  {}{}
}%
\makeatother
\begin{document}

\preprint{AIP/123-QED}

\title{Effect of Charge and Solvation Shell on Non-Radiative Decay Processes in s-Block Cationic Metal Ion Water Clusters.}
\author{Ravi Kumar}
\affiliation{Academy of Scientific and Innovative Research (AcSIR), Ghaziabad-201002, India}
 \altaffiliation[Also at ]{Physical and Materials Chemistry Division, CSIR-National Chemical Laboratory, Dr. Homi Bhabha Road, Pune, 411008, India}
 
\author{Aryya Ghosh}%
 \email{aryya.ghosh@ashoka.edu.in}
\affiliation{Department of Chemistry, Ashoka University, Sonipat, Haryana,131029 India}

\author{Nayana Vaval}
\email{np.vaval@ncl.res.in}
\affiliation{Academy of Scientific and Innovative Research (AcSIR), Ghaziabad-201002, India}
 \altaffiliation[Also at ]{Physical and Materials Chemistry Division, CSIR-National Chemical Laboratory, Dr. Homi Bhabha Road, Pune, 411008, India}

\date{\today}

\begin{abstract}
A molecular cluster's inner valence ionized state undergoes autoionization, which is nonlocal by nature. 
In a molecular system, when the inner valence's ionization potential (IP) is higher than the double ionization energy (DIP), 
it is energetically favorable for the initially ionized system to emit a secondary electron and reach a final state which is lower in energy. 
This relaxation usually happens via intermolecular coulombic decay (ICD) or electron transfer-mediated decay (ETMD). 
We have choosen the Na$^+$-(H$_2$O)$_{n=1-5}$ and Mg$^{2+}$-(H$_2$O)$_{m=1-5}$ cluster as the test systems. 
These systems are also found in the human body, which makes this study important. 
We have calculated the IP, DIP values, and the lifetime of Na-2s and Mg-2s temporary bound states (TBSs) 
in these clusters to study the effect of solvation on IP, DIP, and the lifetime of Na-2s and Mg-2s TBSs. 
We observe a considerable increase (96\%) in the lifetime of the Na-2s TBS in the second solvated shell structure in Na$^+$-(H$_ 2$O)$_{n=2}$ compared to the first solvated one. 
However, the increase in the lifetime of the Mg-2s state in the second solvation shell is only 33\%. 
We have revealed the different factors that affect the lifetime of TBSs and which type of decay process (ICD or ETMD) is dominant. 
We have shown how the charge of metal ions and increased water molecules affect the decay rate. 
We have shown that the decay of Mg-2p is also possible in all magnesium-water clusters, but it is not valid for the decay of Na-2p.
\end{abstract}

\maketitle

%
\section{Introduction}

In nature, we find various metal ions that act as catalysts and are necessary for enzymatic activity. Few metal ions play an essential role in many biological processes. Sodium and Magnesium ions are such ions, out of a few (i.e., Na, Mg, K, Ca, Zn, Fe, Cu, etc.), that are essential for the human body. Studies \cite{300enzym,mg-role1,mg-role2} have shown that magnesium ion alone has been involved in more than 300 enzymatic systems as a cofactor. For example, magnesium is required in the human body for making proteins, maintaining the health of muscles and neurons, controlling blood glucose levels, and structural development of bone. Additionally, magnesium aids in the active movement of calcium and potassium ions across cell membranes, which is necessary for the conduction of nerve impulses, the contraction of muscles, and a regular heartbeat. Magnesium also helps in adenosine tri-phosphate (ATP) production (the energy currency of the human body). Three sodium and two potassium ion use this energy currency to move in and out of the cells. Na$^+$ ion is one of the main elements of the sodium-potassium pump in the human body. Na$^+$ ion also plays an important role in neural signaling \cite{brainsignals} and preventing brain disease.\cite{braindis} Studying non-radiative decay processes in microsolvated clusters will provide insight into the chemistry involved in radiation damage in the human body. 

Most of the human body's chemical and biological reactions occur in the liquid phase. But studying the reactions in the liquid phase is more challenging than in the gaseous phase because of the large number of solvent molecules and weak interactions (solvent-solute and solvent-solvent weak interactions) between these large numbers of molecules in the liquid phase. But, understanding these weak interactions is essential for more accurate results. Thus one can use the concept of micro-solvation, where calculations are performed in the gaseous phase using a few molecules of solvent (at least up to 1st solvation shell). Micro-solvation includes weak interaction, is not too computationally expensive, and provides good-quality results.

\begin{figure}
    \centering
\includegraphics[width=0.5\textwidth]{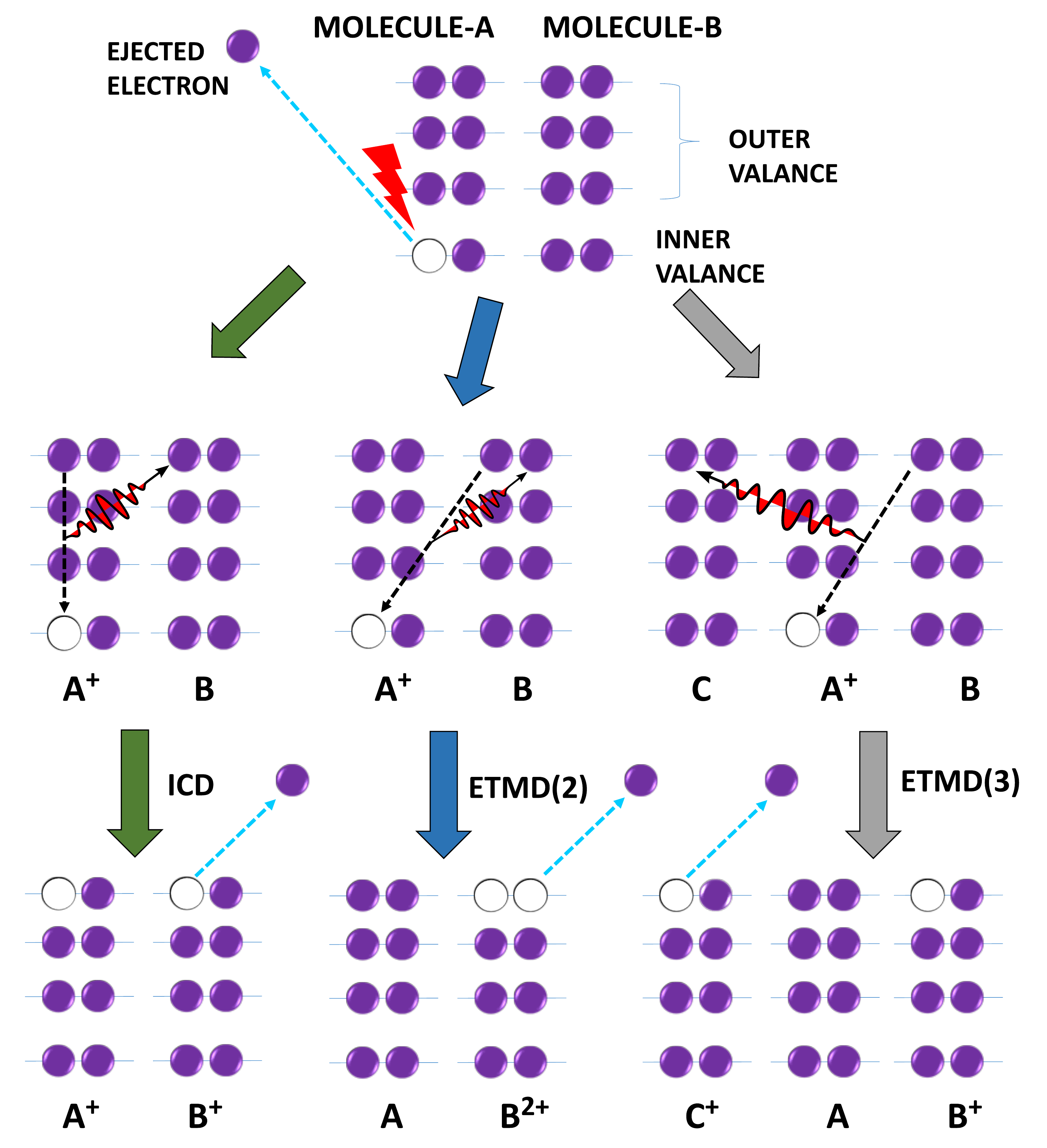}
    \caption{Different types of non-radiative decay processes are represented visually. 
    A vacancy is created in the inner valence shell (mainly in 2s orbitals) due to high-energy electron or X-rays bombardment. (a) In ICD, the vacancy is filled by a valence electron of molecule A and the released energy will knock out molecule B's valance electron. A$^+$B$^+$ type molecular state will be the final state. (b) In ETMD(2), an outer valence electron from the nearby molecule B fills an inner valence vacancy. The virtual photon will eject another valence electron from molecule B, resulting in the final molecular state of  AB$^{2+}$ type. (c) In ETMD(3) process, a neighboring molecule B's outer valence electron fills the hole in molecule A, and the virtual photon ejects a second electron from molecule C. AB$^+$C$^+$ type molecular state will be the final state. In ETMD(2) and ETMD(3), 2 and 3 are the number of molecules involved in the ETMD process.}
    \label{fig:effects}
\end{figure}

Our chosen study system is so versatile that many studies \cite{diff-cation,jctc-ravi,exp-na-h2o6,na_geom,auger_geom_mg_na,na-geom-thrm} have
been conducted. Most of these investigations target the spectra, thermal stability, and global minimum for the ground state structure of the clusters. 
Micro-solvated metal ions can serve as a valuable model for studying solution chemistry at the molecular level. 
These studies are essential for solvation chemistry, electron transfer, charge-induced reactivity, and other properties. 
But only a few studies \cite{auger_geom_mg_na,microsolv} have been done on the non-radiative decay process
using a micro-solvated cationic metal ion water system. Before discussing those systems,
let's understand non-radiative decay processes. As the name implies, there will be no emission of radiation or photon as the
end result. A lifetime of non-radiative decay processes is of a few femtoseconds. Both factors make non-radiative decay 
processes hard to detect. Intermolecular or interatomic coulombic decay (ICD), Electron transfer-mediated decay (ETMD) 
are examples of non-radiative decay processes.In ICD, the initial vacancy of the inner valence subshell is filled by an electron 
from the outer valence shell of the same molecule (from molecule-A). 
The excess energy knocks out an electron from the nearby molecule's valence shell (from molecule B). 
A$^+$B$^+$ type of final state will be formed in ICD. Since there would be a +1 charge on each molecule, 
therefore there would be a Coulomb explosion that would lead molecules away from each other. 
On the contrary, in ETMD, an adjacent molecule's electron (electron from molecule B) fills the initial vacancy. 
The energy released from this process knocks out an electron from the valance shell of neighbor molecules.  
For more details on non-radiative decay, you can see figure \ref{fig:effects}. 

Recently, various factors \cite{review,jctc-ravi,proton,Hemmerich} that affect the non-radiative decay process have been studied, 
for example, the effect of protonation and deprotonation, \cite{proton,Kryzhevoi} bond length,\cite{Ghosh14b} polarization, different molecular environment effect, \cite{jctc-ravi} etc. 
Stumph et al. \cite{lifetime-mg-na} studied the decay width of the  Na$^+$-(H$_2$O)$_n$; n=1,4  and Mg$^{2+}$-(H$_2$O)$_m$; m=1,6 clusters as a
function of metal-oxygen distance and also as the number of nearest neighbors. Their study starts with the optimized geometry of the
cluster with the number of highest water molecules. i.e., Na$^+$-(H$_2$O)$_4$ and  Mg$^{2+}$-(H$_2$O)$_6$ and then take off one water molecule at a time to get new geometry. The water molecule is removed from the cluster in such a way that the number of cis-pairs remains as high as possible. It means they did not use the optimized geometries for Na$^+$-(H$_2$O)$_{1-3}$ and Mg$^{2+}$-(H$_2$O)$_{1-5}$ clusters. 
We know that decay width (or lifetime) highly depends on the geometry of any cluster and even on bond length.\cite{jctc-ravi}
Adding a new water molecule near Na$^+$-H$_2$O and Mg$^{2+}$-H$_2$O ion changes the previous Mg$^{2+}$-O/Na$^+$-O bond distance.
Thus use of optimized structure and unoptimized structure will lead to entirely different results. In our study, we optimized the
geometry of each cluster and then calculated the decay width at the optimized geometry.

In this paper, we have studied the effect of solvation shell on ionization potential (IP) and double ionization potentials (DIP) of Mg$^{2+}$-(H$_2$O)$_{1-5}$ and  Na$^+$-(H$_2$O)$_{1-5}$.
We have investigated the decay mechanism of the inner valence state and analyzed the possible role of the charge of metal ion, cluster geometry, and 
the effect of the solvation shell on the decay process. 
Our main goal is to understand how the solvent molecules in different solvation shells play an essential role in non-radiative decay processes (ICD/ETMD).

\section{Computational Methods}

\subsection{FANO approach for decay width}

The metastable states generated via photoionization can decay through Auger decay, ICD, and ETMD. 
The decay happens through two-electron autoionization mechanisms; hence, these decay processes can belong to the Feshbach-type resonance states. 
The characteristic of these metastable states is governed by the system's decay width $\Gamma$. 
According to Fano-Feshbach's theory, the resonance state can be described as a combination of bound and continuum states. 
The coupling between the bound and continuum part of the resonance state gives us the decay width. The decay width $\Gamma$ in this approach is provided by

\begin{eqnarray}
\Gamma= 2 \pi \sum_{f} |\langle \phi |H| \chi_{f,\epsilon } \rangle |^{2}
\end{eqnarray}

where H  is the total electronic hamiltonian, f indicates open decay channels that belong to the continuum subspace, and  
$\epsilon$ is the asymptotic kinetic energy of the ICD/ETMD electron.
To construct the bound and continuum parts of the resonance state, we have divided the Hilbert space into the Q subspace for bound configurations and 
the P subspace for continuum configurations. Where P and Q subspace would have to fulfill  the following conditions, which are P+Q=1 and P*Q=0.
We have obtained the bound part of the resonance state through the diagonalization of H projected on the Q subspace. 

\begin{eqnarray}
QHQ |\phi \rangle = E_{b} |\phi \rangle
\end{eqnarray}

We have obtained the continuum part of the resonance state through the diagonalization of H projected on the P subspace. 

\begin{eqnarray}
PHP | \chi_{f,\epsilon } \rangle = E_{f} |\chi_{f,\epsilon } \rangle
\end{eqnarray}

The continuum states we obtained using equation-3 are not true continuum because the L$^2$ basis set has been used in our calculations. 
They show wrong normalization and asymptotic behavior. Therefore, they are pseudo-continuum in nature. 
To obtain the correct normalization and accurate value of decay width, we have employed the Stieltjes imaging technique.\cite{reinhart}   

For the construction of H, we have used the extended second-order algebraic diagrammatic construction (ADC(2)X) scheme of the Green's function 
(defined within the space spanned by the one-hole (1h) and two-hole one particle (2h1p) configurations). 
In ADC(2)X approach,\cite{fano,fanoadc,fanova} the coupling between the 1h configurations is treated as the second order of perturbation theory. 
However, the coupling between 1h and 2h1p configurations and the coupling between 2h1p configurations is treated to first-order perturbation theory. 
We considered the 2h1p configurations to construct the P subspace, representing the final double-ionized state with an outgoing free electron. 
In this approach, we can get the total and partial decay width. For details of Fano-ADC method, please see the reference. \cite{fano,fanoadc,fanova}
  
\section{Results and Discussion}

\subsection{Cluster geometries and basis set}

\begin{figure}
    \centering
\includegraphics[width=0.5\textwidth]{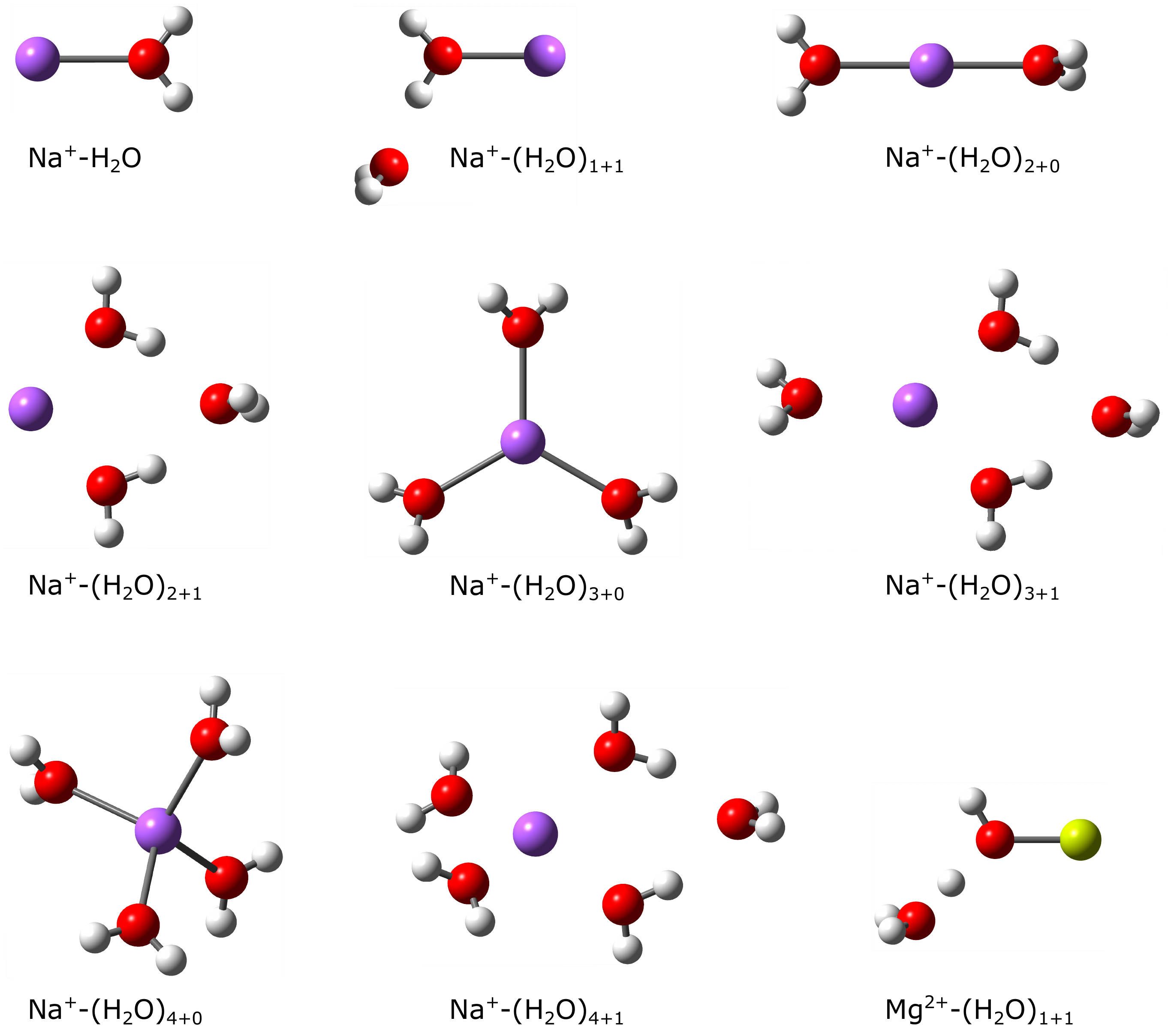}
    \caption{Geometries of the first and second solvation shells of Na$^+$-(H$_2$O)$_{n=1-5}$ and Mg$^{2+}$-(H$_2$O)$_{1+1}$ clusters. Atom colour code Yellow: Magnesium, Violet: Sodium, Red: Oxygen, and White: Hydrogen.}
    \label{fig:sodium-water}
\end{figure}

Geometries of  Na$^+$-(H$_2$O)$_n$  and Mg$^{2+}$-(H$_2$O)$_n$ clusters were optimized using the Gaussian09 software package \cite{G09} with B3LYP \cite{b3lyp, B3, LYP, VWN} functional and 6-311++g(2d,p) basis set. \cite{O2H6311g}
Grimme's GD3 dispersion correction has been used while optimizing geometries. 
Since we are studying the effect of the solvation shell through this article, we must learn a little about the solvation shell.
The number of water molecules (solvent molecules) around a solute (in this case, a metal ion) is known as the solute's solvation shell. 
The solute's solvation shell can be divided into the layer. 
The difference in the first solvation shell structures and the second solvation shell structures are related to the water's position in the given cluster. 
In the first solvation shell structure, all the water molecules are directly connected to the metal ion (according to our system). 
While in the second solvation, all the water molecules are not connected directly to the metal ion. Let us use our test systems to gain a general understanding of this. 
In Na$^+$-(H$_2$O)$_n$ and Mg$^{2+}$-(H$_2$O)$_n$, n = p+q where n is the total number of water molecules present in any metal-water cluster. 
P is the total number of water molecules in the first solvation shell that is connected to the metal ion directly, and q is the total number of water molecules in the second solvation shell that is connected to the water molecules in the first solvation shell directly (indirectly connected to the metal ion). 
For example, Na$^+$-(H$_2$O)$_{2+0}$ is the first solvated shell geometry of Na$^+$-(H$_2$O)$_2$. Here both water molecules are in direct contact with Na$^+$ ion. Thus they are present in the first solvation shell of the Na$^+$ ion. 
While Na$^+$-(H$_2$O)$_{1+1}$ is the second solvated shell geometry of Na$^+$-(H$_2$O)$_2$, where one water molecule is in direct contact with Na$^+$ (present in the first solvation shell) and another water molecule is directly connected to the first water molecule.
It means the second water molecule is indirectly connected to Na$^+$. Thus the second water molecule is in the second solvation shell. See figure \ref{fig:sodium-water}  for a better understanding. 
Similarly, Na$^+$-(H$_2$O)$_{3+0}$ and Na$^+$-(H$_2$O)$_{2+1}$ are the first and second solvation geometries for n=3, respectively. 
Although there are numerous structural alternatives \cite{na_geom} for the first and second solvation shell structures for n$\ge$3, we are just considering the lowest energy structure in the first and second solvation shell structures.

The number of water molecules in the first solvation shell of  Na$^+$ ion is reported differently by different articles. 
You can find a different number of the water molecules in the first solvation shell of Na$^+$ ion in various studies through reference \cite{na_geom} and references within it. 
The coordination number of Na$^+$ ion in the gaseous phase is close to 4, while in bulk, it is between 5 and 6. 
This means that in the gaseous phase, the first solvation shell structure for n$\le$4 will be the lowest minimum structure, whereas the second solvation shell structure will be the lowest energy structure for n$\ge$5 in Na$^+$-(H$_2$O)$_n$. 
As for Mg$^{2+}$-(H$_2$O)$_n$ clusters, the coordination number is close to 6. 
The Mg$^{2+}$'s coordination number in Mg$^{2+}$-(H$_2$O)$_n$ tells that the first solvation shell structure will be the lowest minimum structure up to n$\le$6. 

 The structure of Na$^+$-(H$_2$O)$_{1+1}$ differs significantly from Mg$^{2+}$-(H$_2$O)$_{1+1}$ even though both have same symmetry. 
We observed that Mg$^{2+}$-(H$_2$O)$_{1+1}$ exist in ionic form (Mg$^{2+}$-HO$^-$...H$_3$O$^+$) than neutral form (Mg$^{2+}$-H$_2$O...H$_2$O). 
Whereas Na$^+$-(H$_2$O)$_{1+1}$ exist as Na$^+$-H$_2$O...H$_2$O (neutral form). 
We speculated that it is due to the higher charge of the Mg$^{2+}$ ion than the Na$^+$ ion. 
Except Na$^+$-(H$_2$O)$_{1+1}$ and Mg$^{2+}$-(H$_2$O)$_{1+1}$ geometries, rest all the geometries of both types of the cluster are similar. 
Even in most cases, they have identical symmetries too. Since the cluster's geometry also includes its symmetry. Let us examine the symmetries of sodium and magnesium water clusters. 
With the exception of p=2, where the symmetry for Na$^+$-(H$_2$O)$_{2+0}$ is D$_2$, the symmetries of Na$^+$-(H$_2$O)$_{p+0}$ and Mg$^{2+}$-(H$_2$O)$_{p+0}$ are identical for p=1-4, which are C$_{2v}$, D$_{2d}$, D$_3$  and S$_{4}$.
Symmetries of Na$^+$-(H$_2$O)$_{p+0}$ for p=5 and 6 are C$_1$ and S$_6$, whereas for Mg$^{2+}$-(H$_2$O)$_{p+0}$ they are C$_{2v}$, and T$_H$.
The second solvated shell structures of sodium and magnesium-water clusters (Na$^+$-(H$_2$O)$_{p+1}$ and Mg$^{2+}$-(H$_2$O)$_{p+1}$) were also found to have the same symmetries. 
The symmetries are C$_{s}$ for p=1, C$_{2v}$ for p=2, C$_{2}$ for p=3 and C$_{2}$ for p=4. 
In the second solvation shell structure for n=6, the 4+2 structure type (p=4, q=2) has lower energy than the 5+1 type structure (p=5, q=1). 
D$_{2d}$ is the symmetry for 4+2 type structure of Na$^+$ and Mg$^{2+}$ water clusters.
Coordinates of Na$^+$-(H$_2$O)$_n$ and Mg$^{2+}$-(H$_2$O)$_n$ cluster's first and second solvation geometries and the counterpoise correction 
(or basis set superposition error (BSSE)) for these structures are given in the supporting information. 

We have already discussed the basis set for geometry optimization. 
In IP and DIP calculation using CAS-SCF, and MS-CASPT2 methods, \cite{cas-pt2,werner,caspt22} we have used the cc-pVTZ basis \cite{ccpvtz} for the hydrogen atom, 
the aug-cc-pVTZ basis \cite{ccpvtz} for the oxygen atom, and the ANO-L basis \cite{ano-L} for Mg$^{2+}$ and Na$^+$ metal ions. 
We have used MOLCAS 8.4 version \cite{molcas} to calculate the IP values using CASSCF and MS-CASPT2 methods.\cite{cas-pt2,caspt22,werner}
We have used the aug-cc-pVTZ basis set for all atoms (H, O, Na$^+$, and Mg$^{2+}$) while computing IP using the EOM-CCSD method. 

We have used the FANO-ADC(2)X approach \cite{fano,fanoadc,fanova} for the decay width calculations. 
Decay widths are very sensitive to the basis set. The accurate calculation of decay width requires a proper description of the continuum wave function.
Therefore, cc-pVTZ basis set plus extra 4s4p4d function is used for oxygen atoms and metal ions (Na$^+$ and Mg$^{2+}$) and 
cc-pVTZ basis set augmented by 1s1p1d functions are used for hydrogen in Na$^+$-(H$_2$O)$_{1-3}$ and Mg$^{2+}$-(H$_2$O)$_{1-3}$ systems. 
Basis set addition is done using Kaufmann-Baumeister-Jungen (KBJ) \cite{kaufmann} continuum-like basis functions approach.
Adding the extra basis function will provide a good description of the pseudo-continuum. 
However, adding an extra basis function also increases the calculation's cost. 
Thus one has to choose the basis set by keeping the cost of calculation and the result's accuracy in mind. 
Therefore, for the decay width calculation of Na$^+$-(H$_2$O)$_{4,5}$ and Mg$^{2+}$-(H$_2$O)$_4$ systems, the cc-pVTZ basis set augmented with 2s2p KBJ continuum like basis functions on Oxygen, cc-pVTZ basis set for hydrogen and the cc-pVTZ basis set plus 3s3p1d KBJ continuum like basis functions on the Na and Mg atom has been used.

\subsection{Solvation shell's effect on IP and DIP of Na$^+$-(H$_2$O)$_{1-5}$ and  Mg$^{2+}$-(H$_2$O)$_{1-5}$ clusters}

It is well known that the core IP is dominated by relaxation, whereas outer valence IP is governed by correlation. 
In the case of inner valence ionization, relaxation and correlation both play an important role. 
In some cases, the interplay of these effects causes Koopamans' approximation \cite{generalKA} to fail. 
The inner valence ionized state in a surrounding has a finite lifetime and decays via ICD or ETMD. 
During these non-radiative decay processes, a doubly ionized state will form.
The generation of a double ionized state creates significant reorganization/relaxation in the system. 
Shake-up states usually accompany the double ionization potentials. Thus, many-body effects are crucial for DIP calculations. 
There are two possibilities within double ionized states. First is the single site double ionized states (ss-DIP), where two holes are confined on a single site (A$^{2+}$B or AB$^{2+}$). 
Second is two site-doubly ionized states (ts-DIP), where two holes are localized on two different sites (A$^+$B$^+$ or AB$^+$C$^+$). 
You can see these two types of two-hole states in figure \ref{fig:effects}. 
The ss-DIP states are usually higher in energy than the ts-DIP because two positive charges repel each other more strongly if they are on the same molecule rather than on two different molecules. 
The ts-DIP states are usually responsible for the non-radiative decay process since they are likely to be lower than the IP of the inner valence state. 
It implies that decay is only possible when the lowest molecular DIP will lower than the intended state's IP. 
The number of possible decay channels can be obtained by counting all the DIPs which are lower than the intended state's IP. 
The IPs (Na$^+$-2s, Na$^+$-2p, Mg$^{2+}$-2s, and Mg$^{2+}$-2p) and the lowest DIPs of Na$^+$-(H$_2$O)$_{1-5}$ and Mg$^{2+}$-(H$_2$O)$_{1-5}$, 
are reported in table \ref{Na-water-IP} and table \ref{Mg-water-IP}, respectively. 
IPs and DIPs were calculated using the complete active space self-consistent field (CASSCF) method, \cite{casscf} 
a variant of the multi-configurational self-consistent field (MSSCF) method in the MOLCAS 8.4 version. \cite{molcas} 
We have applied the MS-CASPT2 method \cite{cas-pt2,caspt22,werner} (the second-order perturbation) to include the electron correlation in IP and DIP calculations. 
The CASSCF and MS-CASPT2 method involves CAS-active space that contains the number of active orbital and active electrons. 
The details of the active space for Na$^+$-(H$_2$O)$_n$ and Mg$^{2+}$-(H$_2$O)$_n$ clusters are described in the supporting information. 
The IP calculations are also done using EOM-CCSD \cite{eomip} method to check the accuracy of IP values obtained from the MS-CASPT2 method.
According to the level of approximation we have taken, all the IP values of the MS-CASPT2 method are in good agreement with IP values of the EOM-CCSD method.
Out of all MS-CASPT2's IP values, IP values of Na-2p and Mg-2s agree with respective EOM-CCSD's IP values very well. 
The IP values from MS-CASPT2 and EOM-CCSD follow a similar trend. The calculation of DIPs using the EOM-CCSD method is too expensive. 
Therefore, we have used MS-CASPT2 to calculate the DIP values, which is a good combination of accuracy and computational cost.
One will need two different solvation shell structures to study the solvation shell effect. 
We will need at least two water molecules to build the first and second solvation shells. 
We have taken the lowest energy structures of the first and second solvation for our study.
For Na$^+$-(H$_2$O)$_{2,3}$ and Mg$^{2+}$-(H$_2$O)$_{2,3}$, we have studied two different geometries which correspond to the first and second solvation. 
For Na$^+$-(H$_2$O)$_4$ and Mg$^{2+}$-(H$_2$O)$_4$, we did not study the second solvation structure 
because it will be similar to the second solvation structure of Na$^+$-(H$_2$O)$_3$ and Mg$^{2+}$-(H$_2$O)$_3$, respectively.

\begin{table*}
\caption{ Ionization potentials of Na-2s and Na-2p in Na$^+$-(H$_2$O)$_n$ n=1,5 clusters using different methods with the lowest double ionization potential 
(DIP). All the values are in eV}
\newcommand{\mc}[3]{\multicolumn{#1}{#2}{#3}}
\begin{center}
\begin{tabular}{lrrrrrrrrrrr}
\hline
  Cluster & \mc{3}{c}{Na(2s)} & \mc{3}{c}{Na(2p)}\\
\cline{2-4} \cline{5-7}\\
  & CAS & MSCAS & EOM  && CAS & MSCAS & EOM  && Lowest  \\
  & SCF & PT2 & CCSD   && SCF & PT2   & CCSD && DIP  \\
\hline
Na$^+$-H$_2$O            & 80.220 &  79.104 &  80.410 &&  45.362 &  45.226 &  45.552 &&  53.607 \\
\hline
Na$^+$-(H$_2$O)$_{2+0}$  & 79.077 &  77.526 &  78.696 &&  44.301 &  43.892 &  43.702 &&  43.793 \\ 
Na$^+$-(H$_2$O)$_{1+1}$  & 80.138 &  78.669 &  ---    &&  45.389 &  45.743 &  ---    &&  40.556 \\
Geometry A$^b$           & 80.272 &  78.932 &   ---   &&  45.485 &  45.563 &  ---    &&  38.343 \\
\hline
Na$^+$-(H$_2$O)$_{3+0}$  & 78.002 &  76.465 &  77.172 &&  43.267 &  42.450 &  42.260 &&  39.345 \\
Na$^+$-(H$_2$O)$_{2+1}$  & 78.805 &  77.553 &  ---    &&  44.056 &  43.784 &  ---    &&  39.890 \\
\hline
Na$^+$-(H$_2$O)$_{4+0}$  & 77.036 &  75.962 &  75.839 &&  42.348 &  42.323 &  41.008 &&  38.267 \\
Na$^+$-(H$_2$O)$_{3+1}$  & 77.751 &  76.418 &  ---    &&  44.056 &  42.715 &  ---    &&  38.353 \\
\hline
Na$^+$-(H$_2$O)$_{4+1}$  & 76.764 &  75.295 &  ---    &&  42.178 &  41.906 &  ---    &&  37.135 \\
\hline
\hline
\end{tabular} 
\\

b: First solvation geometrical arrangement with O-Na bond distances of second solvation structure. 
Note: CASPT2 mean MS-CASPT2
\end{center}
\label{Na-water-IP}
\end{table*}

From tables \ref{Na-water-IP} and \ref{Mg-water-IP}, we have the following observations and will try to understand the reasons behind them.
First, we observed that the IP values decrease as n increases in the first solvation shell for Na$^+$-(H$_2$O)$_{n=1-5}$. 
The decrease in effective nuclear charge is to blame for this.
The Na atom has a +1 charge, which decreases as the water molecules increase in the first solvation shell of Na$^+$-(H$_2$O)$_{n=1-5}$. 
As the water molecule increases, the charge transfer from the water molecule to the sodium ion also increases. 
This charge transfer makes the system stable by diffusing the charge of sodium over the water molecules.
NBO analysis supports this hypothesis. NBO analysis reveals a charge transfer between the lone pair of oxygen and the 3s orbital of sodium. 
This interaction stabilizes the system by 1.20 kcal/mol, 1.71$\times$2=3.42 kcal/mol, 5.60$\times$3=16.80 kcal/mol, and 7.68$\times$4=30.72 kcal/mol for n=1 to n=4 in Na$^+$-(H$_2$O)$_n$, respectively. 
We observe an increase in charge transfer as n increases, which leaves sodium and magnesium ions with a less positive charge. 
It indicates that the effective nuclear charge (ENC) experienced by sodium's electrons in Na$^+$-H$_2$O is significantly higher than the ENC experienced by sodium's electrons in Na$^+$-(H$_2$O)$_4$.
Low ENC results in low IP since we need less energy to eject an electron.
As a result, the IP value decreases when the number of water molecules in Na$^+$-(H$_2$O)$_{n=1-5}$ cluster's first solvation shell increases.
A similar trend is observed for Mg$^{2+}$-(H$_2$O)$_{n=1-5}$ clusters. 
Since the Mg atom has a +2 charge and the O-Mg bond distance is smaller than the O-Na bond distance, it promotes the charge transfer even more.
Here the charge transfer stabilizes the system by 9.57 kcal/mol, 16.56$\times$2=33.12 kcal/mol, 22.04$\times$3=66.12 kcal/mol, and 26.44$\times$4=105.76 kcal/mol for n=1 to n=4 in Mg$^{2+}$-(H$_2$O)$_n$. 
Because of the higher charge on Mg than Na, the ENC felt by magnesium's electrons is higher than electrons of sodium. 
As a result, we found that Mg-2s and Mg-2p's IP values are higher than Na-2s and Na-2p's IP values, respectively.

\begin{table*}[ht]
\caption{Ionization potentials of Mg-2s and Mg-2p in Mg$^{2+}$-(H$_2$O)$_n$ n=1,5 clusters using different methods with the lowest double ionization potential 
(DIP) of the cluster. All the values are in eV}
\newcommand{\mc}[3]{\multicolumn{#1}{#2}{#3}}
\begin{center}
\begin{tabular}{lrrrrrrrrrrr}
\hline
  Cluster & \mc{3}{c}{Mg(2s)} && \mc{3}{c}{Mg(2p)}\\
\cline{2-4} \cline{5-7}\\
  & CAS & MSCAS & EOM-  && CAS & MSCAS & EOM- && Lowest  \\
  & SCF & PT2   & CCSD  && SCF & PT2   & CCSD && DIP  \\
\hline
Mg$^{2+}$-H$_2$O            & 116.818 &  115.512 &  116.437 &&  76.369 &  76.008 &  75.947 &&  69.416\\
\hline
Mg$^{2+}$-(H$_2$O)$_{2+0}$  & 114.124 &  113.036 &  113.117 &&  73.715 &  73.389 &  72.437 &&  58.940\\
Mg$^{2+}$-(H$_2$O)$_{1+1}$  & 115.131 &  113.226 &  ---     &&  74.886 &  73.879 &  ---    &&  53.525\\
\hline
Mg$^{2+}$-(H$_2$O)$_{3+0}$  & 112.029 &  109.798 &  110.587 &&  71.702 &  70.858 &  69.987 &&  53.361\\
Mg$^{2+}$-(H$_2$O)$_{2+1}$  & 113.661 &  111.920 &  ---     &&  73.334 &  72.709 &  ---    &&  55.211\\
\hline
Mg$^{2+}$-(H$_2$O)$_{4+0}$  & 110.318 &  107.205 &  107.975 &&  70.050 &  69.264 &  ---    &&  51.320\\
Mg$^{2+}$-(H$_2$O)$_{3+1}$  & 111.517 &  109.268 &  ---     &&  71.212 &  70.440 &  ---    &&  50.224\\
\hline
Mg$^{2+}$-(H$_2$O)$_{5+0}$  & 109.005  & 102.529 &  107.312 &&  68.766 &  68.334 &  ---    &&  50.049\\
Mg$^{2+}$-(H$_2$O)$_{4+1}$  & 109.825  & 107.597 &  107.701 &&  69.562 &  68.952 &  ---    &&  48.730 \\
\hline
\end{tabular}
\end{center}
\label{Mg-water-IP}
\end{table*}

Second, we have observed that the IP values are higher for the second solvation structures than the first solvation structure of Na$^+$-(H$_2$O)$_{2-3}$ and Mg$^{2+}$-(H$_2$O)$_{2-3}$.
Let us understand the reason using Na$^+$-(H$_2$O)$_2$ cluster. 
Two water molecules are directly connected to the metal ion in the first solvation shell (Na$^+$-(H$_2$O)$_{2+0}$), while only one is in the second solvation shell (Na$^+$-(H$_2$O)$_{1+1}$).
As a result, in the first solvation shell, more charge transfer will occur between the oxygen's lone pair electrons and the sodium's 3s orbital than in the second solvation shell.
Therefore, a metal ion's positive charge diffuses across two molecules in the first solvation shell, while in the second solvation shell, it diffuses over just one.
As a result, the electrons of Na and Mg experience a larger ENC in their second solvation shell than in their first. 
Hence, the second solvation shell structures of  Na$^+$-(H$_2$O)$_{2-3}$ and Mg$^{2+}$-(H$_2$O)$_{2-3}$ have higher IP values than the first.
Third, we observed that the lowest DIP values decrease as n increases. Once more, ENC is the cause. Let's understand how.
All the sodium water clusters have +1 charges initially. The two-hole final states will be formed with +3 charges on them.
This +3 charge will be on one sodium atom and one water molecule (two subunits) in Na$^+$-H$_2$O.
In Na$^+$-(H$_2$O)$_5$, this +3 charge will be shared by one sodium atom and the five water molecules (six subunits).
Therefore cluster electrons will experience a drop in ENC as cluster size rises. Thus, DIP values will consequently decrease as cluster size increases.

Fourth, we have observed that the lowest DIP value is not following any trend if we compare the lowest DIP values between the first and second solvation shells in Na$^+$-(H$_2$O)$_2$ and Na$^+$-(H$_2$O)$_3$.
On moving from the first to second solvation shell in Na$^+$-(H$_2$O)$_2$, the lowest DIP value decreases while it increases in Na$^+$-(H$_2$O)$_3$. 
Therefore, we cannot explain the result based on ENC. 
Since the number of water molecules in the first and second solvation structures are also the same, the number of water molecules will also not be the reason. 
Consequently, the cause will depend only on the different positions of the water molecules (geometry). Let's understand the possible connection between them.
SCF calculation shows that the Na-2s and Na-2p's orbital electrons have greater energies than the O-2s and O-2p's orbital electrons.
Therefore, the two-hole state corresponding to the lowest DIP will be on the water molecules, not the sodium atom. 
As soon as two-sited doubly-ionized states form, it will create a lot of strong coulomb repulsion between two water molecules in Na$^+$-(H$_2$O)$_{1+1}$ than Na$^+$-(H$_2$O)$_{2+0}$. 
That is because of the shorter distance (H-H and O-O) between two water molecules in Na$^+$-(H$_2$O)$_{1+1}$ than Na$^+$-(H$_2$O)$_{2+0}$ (for geometry see figure \ref{fig:sodium-water}). 
We have mentioned the H-H bond distance because if a water molecule gets a positive charge, then most of this positive charge will be on two hydrogens of the water molecule than oxygen. 
After all, oxygen is a more electronegative atom than hydrogen.
Since oxygen is a more electronegative atom than hydrogen, if a water molecule receives a positive charge, 
most of this positive charge will be on two hydrogens rather than oxygen, which is why we have mentioned the H-H bond distance.
We have observed that the lowest DIP value is lesser in Na$^+$-(H$_2$O)$_{1+1}$ than in Na$^+$-(H$_2$O)$_{2+0}$ 
because repulsion is dominant between the molecules in the double-ionized state of Na$^+$-(H$_2$O)$_{1+1}$.
We are referring to the H-H bond distance as the distance between two hydrogens from two different water molecules. 
You can ask a question here. What about the repulsion between two hydrogens from the same water molecules? Will it not contribute? 
It will contribute, but this repulsion is present in both the first and second solvation shells. 
Therefore, we can not explain why the lowest DIP of the first solvation shell is higher than the second solvation shell. 
An explanation should be based on things that differentiate first and second solvation shells. 
Therefore geometry is the reason. In other words, the different position of water is the reason. 

In Na$^+$-(H$_2$O)$_3$, the lowest DIP of the second solvation shell is higher than that of the first solvation one.
The reason is, again, the difference in geometry.
The DIP analysis confirmed that the formation of two holes state is on two water molecules that are in direct contact with the Na$^+$ ion in Na$^+$-(H$_2$O)$_{2+1}$.
As two hole state is formed, most of the positive charge of water molecules will be shifted to their hydrogen atoms. The same will happen to the water molecules in Na$^+$-(H$_2$O)$_{3+0}$.
However, the charge of hydrogen atoms in Na$^+$-(H$_2$O)$_{2+1}$ is stabilized by the lone pair electron density of the third water molecule present in the second solvation shell.
NBO analysis shows that O-H interaction stabilizes the system by 8.28$\times$2=16.56 kcal/mol.
This kind of interaction is impossible in Na$^+$-(H$_2$O)$_{3+0}$ because no water molecule is present in the second solvation shell.
That is why we have observed an increase in the lowest DIP value in Na$^+$-(H$_2$O)$_{2+1}$ than Na$^+$-(H$_2$O)$_{3+0}$.

\begin{table*}[ht]
\caption{Lifetime (Decay width) of the Na-2s TBS in Na$^+$-(H$_2$O)$_n$; n=1,5 clusters}
\newcommand{\mc}[3]{\multicolumn{#1}{#2}{#3}}
\begin{center}
\begin{tabular}{lrrrrrrrrrrr}
\hline
  Cluster                & Solvation  & O-Na bond & \mc{2}{c}{Na-2s} \\
\cline{4-5} \\
                         & & length ({in \AA}) & decay width & Lifetime    \\
                         & &            &  $\Gamma$ (in meV) & $\tau$ (in fs)  \\
\hline
Na$^+$-H$_2$O            &  First &  2.24        &  113.97 &  5.78 \\       
Stumph {\it et al.}      &  First &  2.30        &  ---    &  7.0$^a$\\
\hline
Na$^+$-(H$_2$O)$_{2+0}$  &  First &  2.26        &  302.89 &  2.17 \\
Geometry-A$^b$           &  First &  2.18, 4.12  &  209.54 &  3.14  \\
Na$^+$-(H$_2$O)$_{1+1}$  & Second &  2.18, 4.12  &  154.66 &  4.25 \\
\hline
Na$^+$-(H$_2$O)$_{3+0}$  &  First &  2.29        &  335.25 &  1.96  \\
Na$^+$-(H$_2$O)$_{2+1}$  & Second &  2.23, 3.88  &  260.37 &  2.53  \\
\hline
Na$^+$-(H$_2$O)$_{4+0}$  &  First &  2.32        &  400.64 &  1.64 \\
Stumph {\it et al.}      &  First &  2.30        &  365.00 &  1.8$^a$\\
Experimental             &        &  ---         &  ---    &  3.1$^c$\\
\hline
Na$^+$-(H$_2$O)$_{4+1}$  & Second & 2.32, 2.30   &  419.73 &  1.57 \\
                         &        & 4.05         &         &       \\
\hline
\end{tabular}
\end{center}
Note: the O-Na Bond lengths are rounded off up to two decimal place.\\
a:Fano-ADC-Stieltjes method used. Na and O cc-pCVTZ+2s2p2d1f KBG basis; H:cc-pVTZ +1s1p1d KBG functions \\
\vspace{.25cm}
b: First solvation geometry with asymmetric placement of Oxygens  \\
\vspace{.25cm}
c:Experimetal value in aqueous solutions of NaCl Ref.\cite{exper} \\
\vspace{.25cm}
\label{Na-decay}
\end{table*}

\begin{table*}[ht]
\caption{Lifetime (Decay width) of the Mg-2s TBS in Mg$^{2+}$-(H$_2$O)$_n$; n=1,4 clusters}
\newcommand{\mc}[3]{\multicolumn{#1}{#2}{#3}}
\begin{center}
\begin{tabular}{lrrrrrrrrrrr}
\hline
  Cluster                & Solvation  & O-Mg bond & \mc{2}{c}{Mg-2s} \\
\cline{4-5} \\
                         & & length ({in \AA}) & decay width & Lifetime    \\
                         & &            &  $\Gamma$ (in meV) & $\tau$ (in fs)  \\
\hline
Mg$^{2+}$-H$_2$O           &  First &  1.93       &  266.50 &  2.47  \\
Stumph {\it et al.}        &  First &  2.08       &  178    &  3.6 \\
\hline
Mg$^{2+}$-(H$_2$O)$_{2+0}$ &  First &  1.95, 1.95 &  508.55 &  1.29  \\
Mg$^{2+}$-(H$_2$O)$_{1+1}$ & Second &  1.84, 3.95 &  382.31 &  1.72  \\
\hline
Mg$^{2+}$-(H$_2$O)$_{3+0}$ &  First &  1.97       &  588.46 &  1.12  \\
Mg$^{2+}$-(H$_2$O)$_{2+1}$ & Second &  1.92, 3.41 &  512.82 &  1.28  \\
\hline
Mg$^{2+}$-(H$_2$O)$_{4+0}$ &  First &  2.00       &  ---    &  ---  \\
Stumph {\it et al.}        &        &  2.08       &  560    &  1.17 \\
\hline
Experimental               &        &             &  ---    &  1.5 \\
\hline
\end{tabular}
\end{center}
Note: the O-Mg Bond lengths are rounded off up to two decimal place.\\
a:Fano-ADC-Stieltjes method used. Mg and O cc-pCVTZ+2s2p2d1f KBG basis; H:cc-pVTZ +1s1p1d KBG functions \\
\vspace{.25cm}
b: First solvation geometry with asymmetric placement of Oxygens  \\
\vspace{.25cm}
c:Experimetal value in aqueous solutions of MgCl$_2$ Ref.\cite{exper} \\
\vspace{.25cm}
\label{Mg-decay}
\end{table*}


\subsection{Solvation shell's effect on lifetimes of Na-2s and Mg-2s TBSs}

As we have already discussed, Mg-2s/Na-2s TBS's decay is only possible when the IP of Mg-2s/Na-2s state is higher than the lowest DIP of the system.
Furthermore, the number of DIPs smaller than the targeted state's IP determines the number of possible decay channels.
The localization of the final two-hole state will decide whether the non-radiative decay process is ICD or ETMD.
We have reported the decay width $\Gamma$  (in meV) and lifetime  $\tau$ (in fs) of Na-2s and Mg-2s TBSs 
in Na$^+$-(H$_2$O)$_{1-5}$ and Mg$^{2+}$-(H$_2$O)$_{1-4}$ clusters in tables \ref{Na-decay} and \ref{Mg-decay}, respectively.
We have compared our results with theoretical calculations of Stumph {\it et al.},\cite{lifetime-mg-na} 
and with the experimental value \cite{exper} of Na-2s and Mg-2s TBS's lifetime in NaCl and MgCl$_2$'s aqueous solutions.
The systems studied by us and Stumph {\it et al.} are the same.
However, they kept the O-Na and O-Mg bond lengths identical in their studies for all the Na$^+$-(H$_2$O)$_n$ and Mg$^{2+}$-(H$_2$O)$_n$ clusters, respectively.
We have used the optimized geometries of these clusters.
The O-Na bond length varies from 2.24 {\AA} to 2.32 {\AA} for n=1 to 4 in Na$^+$-(H$_2$O)$_{n}$, while Mg-O bond length varies from 1.93 {\AA} to 2.00 {\AA} in our test systems.

We know that bond length also affects the lifetime of a TBS, and a shorter bond length gives faster decay.
For example, the lifetime of Na-2s TBS in Na$^+$-H$_2$O is lower in our calculation (5.78 fs) than the Stumph {\it et al.} results (7 fs).
The difference in results is about 20\%.
This difference is because the basis set and the bond length are different in both studies.
The O-Na bond lengths are 2.30 {\AA} and 2.23 {\AA} for Na$^+$-H$_2$O in Stumph {\it et al.} and our's studies, respectively.
The O-Na bond length remains the same for all the n values in Na$^+$-(H$_2$O)$_n$ clusters in the calculations of Stumph {\it et al.}.

Now we will discuss the changes in decay width of Na-2s and Mg-2s (or lifetime) and explain the difference in Na-2s and Mg-2s TBS's lifetimes in their respective clusters.
We observe that the decay width increases as water molecules increase.
However, we have also noticed that the change in the O-Na and O-Mg bond distances are not much on moving to higher n in the Na$^+$-(H$_2$O)$_n$ and Mg$^{2+}$-(H$_2$O)$_n$ cluster, respectively.
Therefore we have concluded that the increased number of water molecules affects the decay width more.
The number of DIPs that are lower than the targeted state's IP increases with the increasing water molecules.
For a large n value, Na-2s and Mg-2s TBSs will have more decay channels to decay.  
Therefore we observed a faster decay (or a shorter lifetime) of Na-2s/Mg-2s TBS with an increase of n in Na$^+$-(H$_2$O)$_n$ and Mg$^{2+}$-(H$_2$O)$_n$.

We have studied two possible structures of dimer and trimer to study the solvation shells effect. 
It has been observed that the lifetime of Na-2s is higher for Na$^+$-(H$_2$O)$_{1+1}$ (4.26 fs) than Na$^+$-(H$_2$O)$_{2+0}$ (2.17 fs). 
Since the number of possible decay channels is almost identical in both cases, the reason can only be explained based on the O-Na bond distance and 
the position of the second water molecule. These two factors are the primary differences between both structures. 
There is one more factor that can affect the Na-2s TBS's lifetime, and that is the type of non-radiative decay process.
It may be possible that ICD would be the dominant channel in Na$^+$-(H$_2$O)$_{1+1}$ while ETMD would be in Na$^+$-(H$_2$O)$_{2+0}$ and vice-versa.   
To know that, we have done the partial decay width calculations. 
The partial decay width calculations reveal that ICD and ETMD decay processes are 1.88 and 2.69 times faster in Na$^+$-(H$_2$O)$_{2+0}$ than Na$^+$-(H$_2$O)$_{1+1}$. 
However, ICD is the dominant decay channel for both geometries. 
Now the question is which affects the  Na-2s TBS's lifetime more: the position of the water molecule or the O-Na bond distance. 
To understand this, we have studied a new geometry of Na$^+$-(H$_2$O)$_2$ cluster will be known as geometry-A. 
Geometry-A has similar water positions as it in Na$^+$-(H$_2$O)$_{2+0}$ where both water molecules are in the first solvation shell (in direct contact with Na$^+$ ion). 
The O-Na bond length of geometry-A is similar to O-Na bond lengths of Na$^+$-(H$_2$O)$_{1+1}$ which are 2.179 {\AA} and 4.121 {\AA}. 
The O-Na bond length is given in table \ref{Na-decay} for Na$^+$-(H$_2$O)$_{1-5}$ clusters.
Comparison of Na-2s TBS's lifetime in geometry-A and Na$^+$-(H$_2$O)$_{2+0}$ will provide the effect of bond length on Na-2s TBS's lifetime 
because the difference in both geometries is of the O-Na bond lengths while the position of water molecules is the same in both cases. 
Whereas difference in Na-2s TBS's lifetime for geometry-A and Na$^+$-(H$_2$O)$_{1+1}$ will provide the effect of position on Na-2s TBS's lifetime 
because the difference in both geometries is of the different water molecular position while the O-Na bond length is the same in both cases.
The Na-2s TBS's lifetimes are 2.17 fs, 4.25 fs, and 3.14 fs in Na$^+$-(H$_2$O)$_{2+0}$, Na$^+$-(H$_2$O)$_{1+1}$ and geometry-A, respectively. 
The differences in Na-2s TBS's lifetimes are 0.97 fs and 1.115 fs in Na$^+$-(H$_2$O)$_{2+0}$, and Na$^+$-(H$_2$O)$_{1+1}$, respectively. 
The basis of these differences is geometry-A's lifetimes of Na-2s TBSs.
This clarifies that the effect of the positions of water molecules (1.11 fs) is slightly higher than the effect of bond length (0.96 fs) on Na-2s TBS's lifetime.
Therefore we have concluded that the effect of the second water's position (or geometric effect) is dominant over the bond length effect.
Besides these, we also observe that the Na-2s/Mg-2s TBS's lifetimes in all the second shell structures are higher than in their respective first solvation shell structures. 
For example, the lifetime of Na-2s TBS is higher in Na$^+$-(H$_2$O)$_{2+1}$ than in Na$^+$-(H$_2$O)$_{3+0}$.
Coincidentally in the first solvation shell structures, the dipole moments of Na/Mg 2s-ionized states and 
the change in the dipole moment of 2s ionized state (with respect to the ground state's dipole moment) are nearly zero. 
While in the second solvation shell structures of all clusters, neither the dipole moments of Na/Mg 2s-ionized states are zero nor the change in dipole moments are zero. 
We have provided dipole moments for all structures in the supporting information. 
We have observed that the dipole moment decreases as the cluster size in the second solvation shell increases. 
It points out that the dipole moment also plays an important role in stabilizing the Na-2s/Mg-2s TBS. 

We have also noticed that the drop in Na-2s TBS's lifetime is small between any two adjacent first-solvated structures except Na$^+$-H$_2$O and Na$^+$-(H$_2$O)$_{2+0}$.
Two factors are responsible for this large drop in Na-2s TBS's lifetime between Na$^+$-H$_2$O and Na$^+$-(H$_2$O)$_{2+0}$. 
The first factor is the number of decay channels. The number of decay channels increases by more than three times in Na$^+$-(H$_2$O)$_{2+0}$ than Na$^+$-H$_2$O, 
while it increases by less than twice for any two adjacent first-solvated structures. 
The dipole moment is the second factor.  
Na$^+$-H$_2$O has the non-zero dipole moment, while no other first-solvation shell structure has the non-zero dipole moment in their ground state and during Na-2s ionization. 
Non-zero dipole moment stabilizes the Na-2s TBS's lifetime in Na$^+$-H$_2$O.
The first factor drop the Na-2s TBS's lifetime in Na$^+$-(H$_2$O)$_{2+0}$ more than it should, while the second factor enhances the  Na-2s TBS's lifetime in Na$^+$-H$_2$O. 
Therefore, the gap in Na-2s TBS's lifetime is large between Na$^+$-(H$_2$O)$_{2+0}$ and Na$^+$-H$_2$O. 
A similar kind of large gap has been observed in Mg-2s TBS's lifetime between Mg$^{2+}$-(H$_2$O)$_{2+0}$ and Mg$^{2+}$-H$_2$O. 
However, the gap is larger in sodium than in magnesium. That is due to the charge effect. 
If we compare the TBS's lifetime values in sodium-water (5.78 fs to 1.96 fs), magnesium-water (2.47 fs to 1.12 fs), and 
neon-water \cite{ghosh2013} clusters (86 fs to 16 fs) for n=1 to 3, we will find that the lifetime values drop faster in the neon-water clusters. 
It means the decay rate is fastest in the neon-water clusters and slowest in the magnesium-water clusters. 
This is due to the presence of metal and charge over it. 
You will reach a similar kind of conclusion by comparing the lifetime of TBS in the neon cluster \cite{vaval2007} (168 fs to 33 fs) and 
our studied clusters (sodium-water cluster (2.47 fs to 1.04 fs) and magnesium-water cluster (5.78 fs to 1.64)) for n=1 to 4.
It is also clear that the decay rate, which is 1/R$^6$ for neon clusters, need not be the same for metal-water clusters. 
Therefore we concluded that the higher the charge on metal ion slower would be the decay, and the decay will not occur with the rate of 1/R$^6$.

In Na$^+$-(H$_2$O)$_4$, the lifetime of Na-2s TBS in our results (1.64 fs) and Stumph et al. (1.80 fs) theoretical results agree well with each other.
However, it is much lesser than the experimental value (3.0 fs). 
The discrepancy between theoretical and experimental values lies within the error bars of the FANO-ADC method. 
We know that the solvation shell number of sodium-ion is close to 4 in the gas phase. 
Therefore a comparison of Na-2s TBS's lifetime in Na$^+$-(H$_2$O)$_{4+1}$ and Na$^+$-(H$_2$O)$_4$ will also tell us about the effect of solvation shell on the decay. 
The Na-2s TBS's lifetime in Na$^+$-(H$_2$O)$_4$ and Na$^+$-(H$_2$O)$_{4+1}$ are 1.64 fs and 1.57 fs, respectively. 
The value is still decreasing on moving out of the first solvation shell sphere because of the increased number of decay channels due to increased water molecules. 
However, the difference in Na-2s TBS's lifetime (0.07 fs) indicates that the effect of the solvation shell is negligible after attaining the first solvation shell sphere.

\section{Conclusions}

In this article, we have studied the effect of solvation shells on ionization potential (IP), double ionization potential (DIP), 
and a lifetime of 2s ionized states using Na$^+$-(H$_2$O)$_{n=1-5}$ and Mg$^{2+}$-(H$_2$O)$_{n=1-5}$ clusters as the test systems. 
We have specifically looked at the impact of several variables on the lifetime of Na-2s/Mg-2s TBS, 
including the number of water molecules present, the solvation shell's effect, and the effect of the metal ion's charge.
The decay width has been calculated using the Fano-ADC(2)X method, whereas the CAS-SCF, MS-CASPT2, and EOM-CCSD methods have been used to report the system's IP. 
The lowest DIPs are reported using the MS-CASPT2 method. 

Conclusions related to IP and DIP are as follows.
First, the lowest DIP and IP values decrease as sodium-water and magnesium-water clusters size increase. 
For example, the  EOM-CCSD IP values of Mg-2s decrease as the size of magnesium-water clusters increases. 
The decrease in IP value with an increase in cluster size is due to a decrease in effective nuclear charge (ENC). 
We have also observed that the IP values would remain decreased continuously even after attaining the first solvation shell. 
However, the decrease in IP will be slight after attaining the first solvation shell. 
Therefore we concluded that only after achieving the first solvation shell would the change in the cluster's IP be insignificant, while it will significantly affect it before that. 
A decrease in the lowest DIP is due to increased water molecules in the cluster, reducing the repulsion between two ionized species. 
The lowest DIP is geometry dependent; therefore, we have not observed any general trend in the lowest DIP values of the first and second solvation shell structures. 
Second, the lowest DIP is less in the second solvation shell of Na$^+$-(H$_2$O)$_2$ than in the first solvation shell of the same cluster. 
This is due to high repulsion between two positively charged water molecules. 
At the same time, the lowest DIP is higher in the first solvation shell of Na$^+$-(H$_2$O)$_3$ than in the second solvation shell of the same cluster. 
The third water molecule present in the second solvation shell stabilizes the charge over the two first solvated water molecules.
Third, the IP of the second solvation is higher than the first solvation shell for Na$^+$-(H$_2$O)$_{n=2,3}$. 
That is due to the direct connection of water molecules in the first solvation shell that stabilizes the positive charge on metal ions, which decreases the ENC.  
Conclusions related to the lifetime of TBS in sodium-water and magnesium-water clusters are as follows.
a.) With the increase in cluster size (or as n increases), the lifetime of TBS decreases. 
That is due to the increase in the number of possible decay channels.
b.) The lifetime of Na-2s/Mg-2s TBS is higher in the second solvated shell structures than in their respective first one.
For example, Na-2s TBS's lifetime is higher in Na$^+$-(H$_2$O)$_{1+1}$ (4.25 fs) than in Na$^+$-(H$_2$O)$_{2+0}$ (2.17 fs). 
That is due to two reasons. (i) there is an increase in one of the O-Na bond lengths in Na$^+$-(H$_2$O)$_{1+1}$, which decreases the rate of decay.
(ii) The change in the position of the second water molecule (asymmetric position) induces the dipole moment in the system. 
This will give rise to polarization in the medium, stabilizing the Na-2s TBS and giving a larger lifetime. 
In other words, Na-2s TBS's lifetime increases by ~96\% in Na$^+$-(H$_2$O)$_{1+1}$ compared to the Na$^+$-(H$_2$O)$_{2+0}$, 
whereas the increase from first to second solvation is only 29\% for trimer. 
This shows that the difference in lifetime between the first and second solvation shells will reduce as water molecules increase. 
As a result, the lifetime of Na-2s/Mg-2s in the metal-water clusters will saturate after completing the first solvation sphere. 
c.) The difference of Na-2s TBS's lifetime (3.61 fs) between Na$^+$-H$_2$O and Na$^+$-(H$_2$O)$_{2+0}$ is higher than the difference of Mg-2s TBS's lifetime (1.18 fs) between Mg$^{2+}$-H$_2$O and Mg$^{2+}$-(H$_2$O)$_{2+0}$. 
Comparing our results (sodium-water and magnesium-water clusters) with the neon-water clusters \cite{ghosh2013} (or neon clusters \cite{vaval2007}), 
we have found that the decay rate is fastest for neon-water clusters (or neon clusters) and slowest for magnesium-water clusters. 
The different decay rates in different clusters are due to the charge effect of metal ions. 
It is also evident that the decay rate for metal-water clusters need not be the same as that for neon clusters, which is 1/R$^6$.
Therefore we concluded that the higher the charge on metal ions slower would be the decay, which will not occur at the rate of 1/R$^6$.
d.) The lifetime of Na-2s TBS is higher than the Mg-2s TBS's lifetime in their corresponding clusters. 
For example, the lifetime of Na-2s TBS in Na$^+$-H$_2$O (5.78 fs) is higher than Mg-2s TBS in Mg$^{2+}$-H$_2$O (2.47 fs). 
The reasons for this are (i) The number of decay channels is more in the magnesium-water clusters than in sodium-water clusters. 
(ii) The metal-oxygen bond lengths are smaller for magnesium-water clusters than sodium-water clusters. 
(iii) The higher charge on magnesium will induce polarization in the surrounding water molecules stabilizing the system. 

\begin{acknowledgments}
RK acknowledges financial support from Council of scientific and Industrial Research (CSIR). Authors acknowledge the computational 
facility at the CSIR-National Chemical Laboratory and Ashoka University. AG acknowledges financial support from SERB (SRG/2022/001115).
\end{acknowledgments}

\section*{Data Availability Statement}
The data that support the findings of this study are available within the article [and its supplementary material].
\section*{References}
\bibliography{solvation-effect}

\end{document}